\documentclass[pra,twocolumn,aps]{revtex4}
\usepackage{amsmath}
\usepackage{wrapfig}
\usepackage[pdftex]{graphicx}
\usepackage{wrapfig}
\usepackage[usenames,pdftex]{color}
\usepackage[ansinew]{inputenc}
\usepackage{calrsfs}
\usepackage{amstext}
\newcommand{\ket}[1]{\smash{|#1\rangle}}              %Ket cmd: \ket{}
\usepackage{gensymb}
\definecolor{rem}{rgb}{1.0,0,0}

\usepackage{comment}
\usepackage{physics}
\usepackage{xcolor}
\usepackage{hyperref}
\newcommand{\bbone}{\text{\usefont{U}{bbold}{m}{n}1}}
\MakeRobust{\bbone}
\usepackage{amssymb}
\usepackage{amstext}

\begin{document}
\title{High entanglement regimes in the Weisskopf-Wigner theory for spontaneous decay}

\author{J. C. C. Capella$^1$, A. Fonseca$^2$, Pablo L. Saldanha$^{3}$, and D. Felinto$^1$}

\affiliation{$^1$Departamento de F\'{\i}sica, Universidade Federal de Pernambuco, 50670-901 Recife, PE - Brazil \\$^3$ Departamento de F\'isica, Universidad Nacional de Colombia, 111321, Bogot\'a, Colombia  \\$^3$ Departamento de F\'{\i}sica, Universidade Federal de Minas Gerais, 30161-970 Belo Horizonte-MG, Brazil}
 

\date{\today}

\begin{abstract}
In this work we review the Weisskopf-Wigner formalism for spontaneous emission considering the spatial modes of light as well as external atomic degrees of freedom which we introduce in the theory by modeling the atom as a wavepacket in momentum space with a given initial uncertainty. We perform a purity calculation in order to quantify the entanglement encoded in the momentum variables of the atom-photon system. Our purity calculations reveal two high entanglement regimes depending on the initial atomic momentum uncertainty: $1)$ the Recoil entanglement regime (which arises in the small momentum uncertainty region) where recoil effects dominate the mechanisms that originate entanglement, and $2)$ the Doppler entanglement regime (in the large momentum uncertainty region) where homogeneous Doppler shifts in the emitted photon's frequency play the fundamental part in the build up of quantum correlations in the system. Physical considerations are made to explain the nature of each entanglement regime as well as provide their respective thresholds. 
% Finally, we briefly investigate the role of entanglement in the distinguishability of two physically different quantum states that arise naturally from the theory, where we note that entanglement in the system leads to a better resolution of the two quantum states.
\end{abstract}

\maketitle
\section{Introduction}

Atom-photon entanglement in spontaneous emission in free space is, presently, a fundamental resource for various quantum information protocols~\cite{Duan2001,Barrett2010,Monroe2013}. Entanglement between single atoms or ions (without cavities) and single photons was experimentally demonstrated by using the internal degrees of freedom of individual emitters to generate polarization entanglement, leading to high entanglement fidelities in the system~\cite{Blinov2004} or violation of Bell inequalities~\cite{Volz2006}. Actually, polarization entanglement between single atoms and photons was implicit already in the early experiments on violation of Bell inequalities in the cascaded decay of atoms~\cite{Freedman1972,Aspect1982}. In recent years, this kind of entanglement has been explored in experiments towards scalable quantum computing or communication with ions~\cite{Inlek2017}. 

Cavities can be used to engineer the spontaneous emission to favor decay into particular modes of the electromagnetic field~\cite{Goy1983}, an effect largely used in the fields of cavity Quantum Electrodynamics~\cite{Miller2005,Walther2006} and its recent development into circuit Quantum Electrodynamics~\cite{Blais2021}. It is also possible to control the spontaneous emission of single photons into particular modes in free space by preparing collective states in large atomic ensembles storing a single atomic excitation~\cite{Chou2004,Matsukevich2006,Chen2006,deOliveira2014}. The storage of such collective atomic excitations explore the hyperfine and Zeeman structures of the atomic ground states, and were used as well to demonstrate polarization entanglement between a single emitted photon and the corresponding collective excitation~\cite{Matsukevich2005,deRiedmatten2006}.

In free space, the basic aspect of the atom-photon entanglement that is critical for quantum information applications is its ability to entangle distant non-interacting material systems, as originally proposed for two individual atoms~\cite{Cabrillo1999,Simon2003} and later expanded for single excitations in atomic ensembles~\cite{Duan2001}. These schemes have been already experimentally demonstrated both to entangle pairs of atoms~\cite{Moehring2007,Slodicka2013} or pairs of atomic ensembles~\cite{Chou2005}. Since then, multipartite entangled states of increasing complexity have been experimentally demonstrated through the engineering of the detection modes of spontaneously emitted photons~\cite{Chou2007,Choi2010,Pu2018}. 

Even though the last twenty years have witnessed this rapid development in the control and application of quantum entanglement between atoms and spontaneously emitted photons, some of its most fundamental features related to strong correlations in the external degrees of freedom of the system remain elusive and largely unexplored. The spontaneous emission from an excited two-level atom is a paradigmatic system for first-principles theories describing the coupling of individual systems to its environment resulting in an irreversible process. In this case, the individual atom interacts with the free-space vacuum modes resulting in its irreversible decay to the ground state~\cite{Dirac1927}. The theoretical analysis of this process was best formulated in the seminal work by Weisskopf and Wigner in 1930~\cite{Weisskopf1930}. The original Weisskopf-Wigner treatment, however, does not include an analysis of the final state of the global system formed by the atom and its emitted photon. It has been long understood that momentum entanglement should exist between the two parts of the system, with the direction of emission being correlated with a corresponding atomic recoil~\cite{Kurtsiefer1997}. However, only more recently the role of the initial atomic wavepackets has been fully appreciated and formally merged into the original Weisskopf-Wigner model~\cite{Rzazewski1992,Stoop1995}. With this, the structure of the underlying multimode atom-photon entanglement in their external degrees of freedom started to be revealed and quantified~\cite{Chan2002,Fedorov2005}. These first analysis already highlighted the existence of regions of parameters with high degrees of quantum correlations. However, it was also clear the complexity of the underlying problem, with an intricate dynamics of the position wavepackets and its corresponding quantum correlations~\cite{Fedorov2005}.

In the present work, we revisit the problem of atom-photon entanglement in the Weisskopf-Wigner theory of spontaneous emission from a two-level atom focusing on its potential as physical resource and on gaining insight on the physical mechanisms behind it. We consider the simplest atomic wavepacket, with a Gaussian profile of minimum uncertainty, and a variable width in momentum space. Our whole analysis is carried out in momentum space, as our goal is to characterize the asymptotic states of the system, in which the system is left after the complete emission of the photon, and not its temporal dynamics. We map then the regimes of high degree of entanglement of the asymptotic state for multiple atoms and atomic transitions and determine the critical parameter values to enter in each regime. In order to quantify the degree of entanglement we use the purity of the reduced bipartite atom-photon system. The rate of enhancement of the corresponding Schmidt number provides insights on the physical origins of the high entanglement. For the regime of recoil entanglement, with small momentum uncertainty for the atom, where the atom recoil in the opposite direction of the photon emission generate atom-photon entanglement, we identify the increase in the number of independent directions of photon emission as the responsible for its increased Schmidt numbers. On the other hand, the regime of Doppler entanglement, with a large momentum uncertainty for the atom, where Doppler shifts on the photon frequency that depends on the atom momentum generates atom-photon entanglement, is dominated by the number of independent spectral regions that can be accessed by the emitted photon. However, depending on the physical parameters associated with the corresponding spectral lines, these two regimes might not be clearly separated or even completely overlap. These mechanisms must be present, even if not completely explored, on the various single-atom traps implemented in the last decades. In the future, we are also particularly interested on how these mechanisms affect the quantum correlations for photons emitted from ensembles of two-level atoms, as recently observed by our group~\cite{Araujo2022,Marinho2023,Marinho2025}.
% There are, however, atomic transitions that present these two mechanisms simultaneously

In what follows, in Section II we revisit the  Weisskopf-Wigner theory for spontaneous emission, obtaining the desired atom-photon quantum state. In Section III we quantify the entanglement present in the atom-photon system by means of a calculation of the purity of the reduced atomic state and observe two high entanglement regimes. In Section IV we provide the physical mechanisms behind the build up of entanglement on each of the found high entanglement regimes as well as the corresponding thresholds. Finally, in Section V we conclude our work and provide perspectives for our findings.
% , in Section V we explore implications of these high entanglement regimes on the discrimination of two paradigmatic quantum states that naturally arise from the theory, and

\section{Weisskopf-Wigner theory for spontaneus emission}

The Weisskopf-Wigner model for spontaneous emission \cite{Weisskopf1930} considers a two-level atom, interacting coherently with the vacuum modes of the electromagnetic field. The Hamiltonian of the system may be written as 

\begin{equation}
\mathbf{H} = \bigg[\frac{\mathbf{p}^2}{2m}+\frac{\hbar\omega_0}{2}\mathbf{\sigma}_z+\sum_{\vec{k},s}\hbar\omega_{\vec{k}}\left(\mathbf{a}_{\vec{k},s}^\dagger\mathbf{a}_{\vec{k},s}+\frac{\bbone}{2}\right)\bigg]+\mathbf{V}_I,
\end{equation}
\noindent 
where $\mathbf{p}$ is the linear momentum operator of the atom and $m$ its mass; $\pm\hbar\omega_0/2$ are the energies associated to the excited and ground states, respectively; $\mathbf{a}_{\vec{k},s}$ denotes the annihilation operator of the $\vec{k}$ mode of the electromagnetic field with angular frequency $\omega_{\vec{k}}$ and polarization $s$; and $\mathbf{V}_I$ corresponds to the interaction Hamiltonian. We employ the electric dipole approximation, and the interaction term reads
\begin{figure} 
  \centering
  \includegraphics[width=0.5\textwidth]{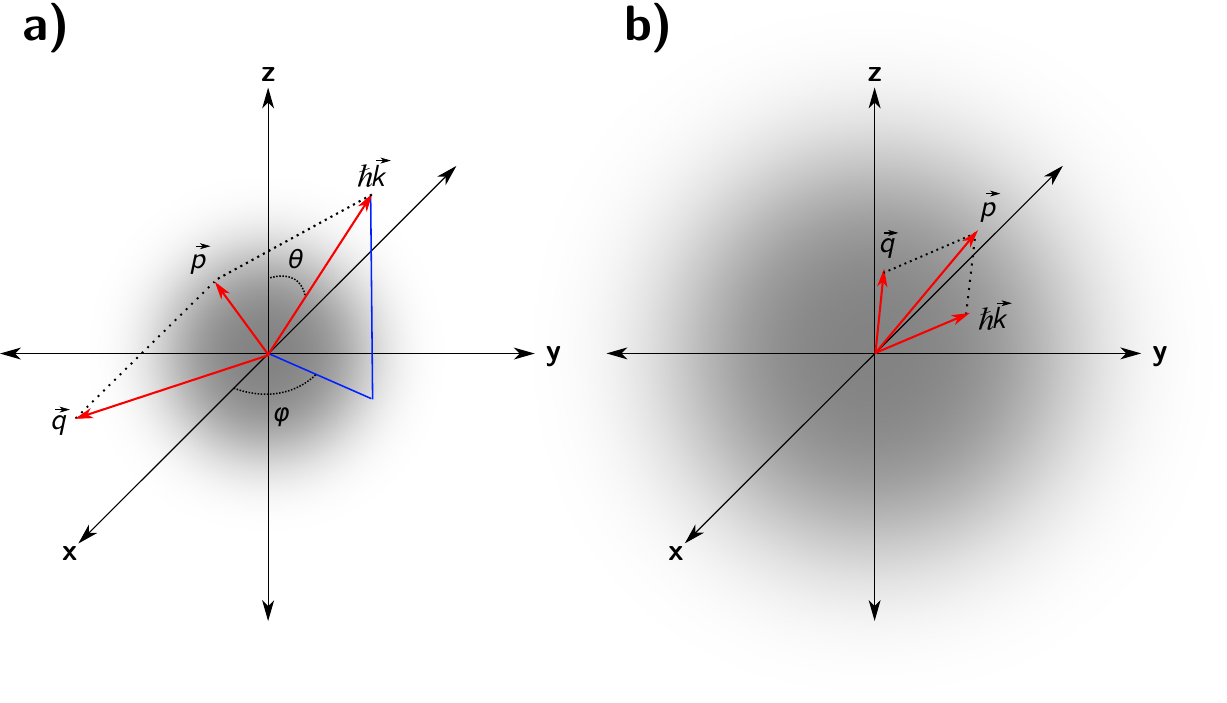}
  \caption{Schematic representation of momentum vectors in the process of spontaneous emission. The shadow regions represent the initial distribution of linear momentum associated with the atom, with a standard deviation $\Delta p$. The vector $\vec{p}$ denotes the initial linear momentum of the atom, while $\hbar\vec{k}$ and $\vec{q}$ represents the final momenta of photon and atom, respectively. In (a) and (b), each situation corresponds to the case where the uncertainty in atomic linear momentum is lower or larger than the momentum of the photon, that is, $\Delta p < \hbar\omega_0/c$ or $\Delta p > \hbar\omega_0/c$, respectively. The angles $\theta$ and $\varphi$ in (a) denote standard spherical coordinates associated with the wave vector $\vec{k}$ of the photon.}
  \label{G10}
\end{figure}

\begin{equation}
 \mathbf{V}_I=\hbar\sum_{\vec{k},s}g_{\vec{k},s}e^{i\vec{k}\cdot\vec{r}}\mathbf{a}_{\vec{k},s}\mathbf{\sigma}_++h.c.,
\end{equation}
\noindent
with  $\mathbf{\sigma}_+$ the raising operator in the internal atomic degrees of freedom, and $g_{\vec{k},s}$ is the coupling constant, given by $g_{\vec{k},s}=-i/\hbar \sqrt{\hbar\omega_k/2\epsilon_0V}e^{-i\vec{k}\cdot\vec{r}_0} ~\vec{t}\cdot\hat{e}_{\vec{k},s}$, with $\vec{t}$ the dipole moment of the transition, $\vec{r}_0$ the position vector of the atom, $\hat{e}_{\vec{k},s}$ is the polarization basis, $\epsilon_0$ is the vacuum permittivity and $V$ the quantisation volume. Consider the system initially prepared in the state

 \begin{equation}
  \ket{\psi_0}=\int \dd^3 p~\varphi(\vec{p}) \ket{e}\otimes\ket{\vec{p}}\otimes\ket{0}_{ph}, 
 \end{equation}

\noindent
where $\ket{e}$ stands for the excited atomic state, $\varphi(\vec{p})$ corresponds to the initial linear momentum distribution of the atom, and the ket $\ket{0}_{ph}$ represents the state of zero excitation in the modes of the electromagnetic field (vacuum). After some steps, it is possible to show that the state of the system evolves as \cite{Rzazewski1992}

\begin{multline}
 \ket{\psi(t)}=\int \dd^3 p ~\varphi(\vec{p})~e^{-\frac{\Gamma}{2}t} \ket{e} \ket{\vec{p}}_{at}\ket{0}_{ph} + \\
 +\sum_{\vec{k},s}\int \dd^3 p ~c_{s}(\vec{p}-\hbar\vec{k},t) \ket{g}_{} \ket{\vec{p}-\hbar\vec{k}}_{at}\ket{1_{\vec{k},s}}_{ph}.
 \label{GenState2}
\end{multline}

\noindent
The first term in Eq. \ref{GenState2} describes the exponential decay of the initial state at a rate $\Gamma = \frac{1}{4\pi \epsilon_0} \frac{4\abs{\vec{t}}^2\omega_{0}^3}{3\hbar c^3} $. The second term accounts for the emission of a photon and subsequent recoil of the atom as it reaches the ground state, with the linear momentum vectors of photon and atom given by $\hbar\vec{k}$ and $\vec{p}-\hbar\vec{k}$, respectively. $c_{s}(\vec{p}-\hbar\vec{k},t)$ is the probability amplitude at time $t$ for the emission of a photon with wavevector $\vec{k}$. The state $\ket{1_{\vec{k},s}}_{ph}$ represents one photon in mode $(\vec{k},s)$ and all other modes in the vacuum. We are interested in the correlations between linear momentum variables in the atom-photon system after the emission process. Then we restrict our analysis to the asymptotic case $\Gamma t \gg 1$, in which it is safe to neglect the first term in Eq. \ref{GenState2} (for an analysis of entanglement in the transient regime, see \cite{Leandro2009}). In this limit, it can be shown that the state is reduced to 

\begin{equation}
 \ket{\psi} \approx \int \dd^3 q~\int \dd^3 k ~ C(\vec{q},\vec{k}) \ket{\vec{q}}_{at} \ket{\vec{k}}_{ph},
 \label{Gen_State_a}
\end{equation}

\noindent
where

\begin{equation}
C(\vec{q},\vec{k})=\frac{-\frac{i}{4\pi}\left(\frac{3\omega_k  c^3\Gamma}{\omega_0^3}\right)^{1/2} \sin\theta~\varphi(\vec{q}+\hbar\vec{k})}{\omega_{\vec{k}}-\omega_{0}+\frac{\left|\vec{q}\right|^2}{2 m\hbar}-\frac{\left|\vec{q}+\hbar\vec{k}\right|^2}{2 m\hbar}+i\frac{\Gamma}{2}},
\label{StateC}
\end{equation}
 
 \noindent
and $\theta$ is the angle between $\vec{k}$ and the $z$ axis, and $\vec{q} = \vec{p}-\hbar\vec{k}$ represents the final atomic momentum (see Fig. \ref{G10}). 

Experimentally it is possible to trap single atoms in a minimum of a given potential, which can be approximated as a harmonic potential. This leads to atomic states corresponding to the ground state of a quantum harmonic oscillator for its external degrees of freedom, that is, a Gaussian wave-packet in momentum \cite{Diedrich1989, Monroe1996, Liebfried2003}. Therefore, we use this experimental picture to assume a Gaussian wave-packet as our initial distribution of atomic momenta. Explicitly, we assume:
\begin{equation}\label{varphi}
 \varphi(\vec{p})=\left(\frac{1}{\pi\Delta p^2}\right)^{3/4}e^{-\frac{\vec{p}^2}{2\Delta p^2}}.
\end{equation}

\section{Quantification of entanglement encoded in the momentum variables}
One of the ways we can assess the presence of entanglement in a pure bipartite quantum state is by evaluating the purity of the reduced state of one of its parts~\cite{Horodecki2009}. The reduced state associated to the atomic subsystem can be written as:
\begin{align}
 \rho_{a} &= \tr_{ph} \left(\dyad{\psi}\right) \nonumber \\
 & = \int \dd \vec{k}~ \dd\vec{p}~ \dd\vec{p'}~C(\vec{p},\vec{k}) C^*(\vec{p'},\vec{k}) \dyad{\vec{p}-\hbar\vec{k}}{\vec{p'}-\hbar\vec{k}}.
\end{align}
Therefore, the purity of the reduced state of the atom reads:
\begin{align}\label{purity1}
p_a & = \tr \left(\rho_a^2\right) \nonumber \\
& = \int \dd \vec{k}~ \dd \vec{k'}~\dd\vec{q}~ \dd\vec{q'}~C(\vec{q}+\hbar\vec{k},\vec{k}) C^*(\vec{q}+\hbar\vec{k'},\vec{k'}) \times \nonumber\\
&\times C(\vec{q'}+\hbar\vec{k'},\vec{k'}) C^*(\vec{q'}+\hbar\vec{k},\vec{k}),
\end{align}
with $C(\vec{q}+\hbar\vec{k},\vec{k})$ given by Eq.~\eqref{StateC}. In this case, $p_a = 1$ indicates a separable state, and $p_a \rightarrow 0$ a highly entangled state.
 
Now we calculate $p_a$ for various physical conditions of the spontaneous emission process, considering different spectral lines for four atoms. Note that most parameters, such as $\Gamma, \omega_0$ and $m$, are fixed for a given spectral line. However, the parameters related to the atomic external degrees of freedom require preparation of the atomic states, as discussed above. We have considered so far, for simplicity, that the atom is positioned at the origin with also zero initial average momentum. The only parameter that is left undetermined is the uncertainty $\Delta p$ in the initial atomic momentum. This is a critical parameter, and we will plot most curves from now on as functions of it.

The uncertainty in $\Delta p$ can be cast as an uncertainty in the kinetic energy of the atom. It is useful then to represent it in terms of a quantity with units of temperature that we call uncertainty temperature $T_u$ and define as
\begin{equation}
    k_b T_u = \frac{\Delta p^2}{2 m}.
\end{equation}
In this way, we can compare directly the  uncertainty in $\Delta p$ with two temperatures representing different regimes of the atomic dynamics, relevant for cold atomic physics experiments in general. The first is the recoil temperature $T_R$, defined as
\begin{equation}
k_b T_R = \frac{\hbar^2 \omega_0^2}{m c^2}\, .
\end{equation}
It represents a scale of variations in atomic momentum of the size of the momentum of a single emitted photon. The other regime is  represented by the Doppler temperature $T_D$, defined as
\begin{equation}
k_b T_D = \frac{\hbar \Gamma}{2}.
\end{equation}
This provides the lowest temperature achievable by the process of Doppler cooling for an ensemble of two-level atoms, indicating the range of kinetic energies above which the atom is more sensitive to the Doppler effect. 

\begin{figure}[htb!]
\centering
\includegraphics[width=0.5\textwidth]{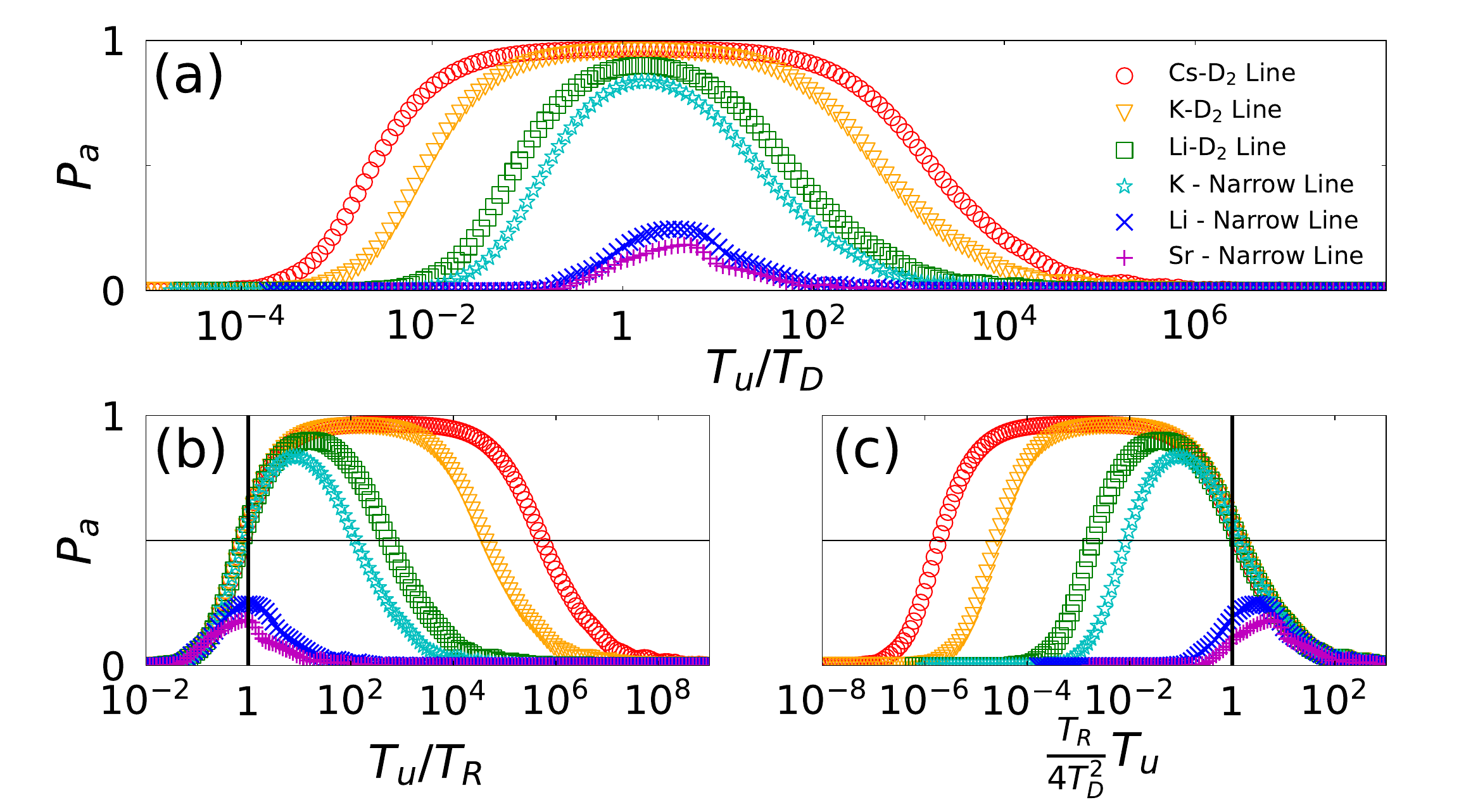}
\caption{Purity of the reduced state for the different atomic spectral lines considered in Table \ref{tab:spectrallines} as a function of (a) $T_u/T_D$, (b) $T_u/T_R$ and (c) $\frac{T_R}{4 T_D^2}T_u$. The vertical solid black lines in (b) at $T_u = T_R$ and in (c) at $T_u = \frac{4T_D^2}{T_R}$, indicate the thresholds for the Recoil and Doppler entanglement regimes, respectively, found  in Section IV. Note that the vertical lines cross most of the graphs at around $p_a \approx 0.5$ (horizontal lines).}

%\hline Li - Narrow Line & \color{red}{inserir} & 27.48 & 8.02 \\

%The solid black vertical lines indicate the cases where $\mathrm{T_u = T_R}$ and $\mathrm{T_u  = T_D}$, respectively on the main plot and on the inset.}
%, while the solid curves give the product of the purities, $P_p$, calculated using the small and large $\Delta p$ approximations given respectively by Eqs. \eqref{smallpapprox} and \eqref{largepapprox}. \color{red}{Refazer rotulos dos eixos, mudando fonte na vertical e passando de energia para temperatura na horizonal. Inserir inset com eixo x em termos de $T_{u}/ T_{D}$.}
\label{purity_fig}
\end{figure}

As expected, the twelve-dimensions integral in Eq.~\eqref{purity1} cannot be put into an analytical form easily. Therefore, we perform a numerical integration (using the quasi-Monte Carlo method of integration embedded into the Mathematica\textregistered\, software version 11.2) to obtain the scattered points in Fig. \ref{purity_fig} for the different spectral lines considered in Table \ref{tab:spectrallines}. This particular choice of spectral lines gives a sizeable range of Doppler and recoil temperatures and in fact, a good range for the quantity $T_D/T_R$, which as we can see from Fig. \ref{purity_fig} yields different behaviours for each of the lines.

One characteristic that is common to all the considered spectral lines is the presence of two high entanglement regimes for small and large $T_u$, illustrated by the decrease in $p_a$ shown in Fig. \ref{purity_fig} on the edges of each line. Some spectral lines, such as $\mathrm{Cs - D_2}$ and $\mathrm{K - D_2}$, separate these high entanglement regimes quite well, with a region of $p_a \approx 1$ in the middle. Others, such as the Sr - Narrow Line, present a high degree of entanglement for the whole range of $\Delta p$ considered.

\begin{table}[h!]
\begin{tabular}{| c | c | c |c|}
\hline 
 Spectral Line & Transition & $T_R(\mu K)$ & $T_D(\mu K)$ \\
\hline
$\mathrm{Cs-D_2}$ Line & $6S_{1/2} \rightarrow 6P_{3/2}$ & 0.20 & 125\\
\hline$\mathrm{K-D_2}$ Line & $4S_{1/2} \rightarrow 4P_{3/2}$ & 0.82 & 144\\
\hline $\mathrm{Li-D_2}$ Line & $2S_{1/2} \rightarrow 2P_{3/2}$ & 6.36 & 140\\
\hline K - Narrow Line & \color{black}{$4 \,^2S_{1/2} \rightarrow 5 \,^2P_{3/2}$} & 1.46 & 28.6\\
\hline Li - Narrow Line & \color{black}{$2 \,^2S_{1/2} \rightarrow 3 \,^2P_{3/2}$} & 15.3 & 18.1 \\
\hline Sr - Narrow Line & \color{black}{$5s^2 \,^1S_0 \rightarrow 5s5p \,^3P_1$} & 0.46 & 0.18 \\
\hline
\end{tabular}
\caption{Recoil and Doppler temperatures for the different spectral lines in Fig.~\ref{purity_fig}. Available in \cite{Ding2016, Steck2003}.}
    \label{tab:spectrallines}
\end{table}
\vspace{-0mm}

%For $T_u \lesssim T_R$, then, $p_a$ goes rapidly to zero as $T_u$ decreases, independently of the parameters of the emission line. Conversely, for $T_u \gtrsim T_D$, $p_a$ decreases rapidly as $T_u$ increases (see inset to Fig.~\ref{purity_fig}). Thus, $T_R$ and $T_D$ determine the limits of the two regions of high entanglement, and the condition $T_D >> T_R$ leads to a clear separation of these regions with the formation of a $p_a \approx 1$ plateau in the middle. 

From Fig. \ref{purity_fig}(a) we observe that the plateau of low entanglement ($p_a \approx 1$) is roughly centered at $T_u = T_D$, while from Fig. \ref{purity_fig}(b) and Fig. \ref{purity_fig}(c) we can note two important parameters: $T_R$ and $\left(\frac{T_D}{T_R}\right)T_D$, respectively, which give a natural order of magnitude for the values of $T_u$ from which we expect high entanglement. Moreover, we note that the behavior of the purity of the reduced state is independent from the atomic line considered for small $T_u$ in Fig. \ref{purity_fig}(b), and for large $T_u$, in Fig. \ref{purity_fig}(c), due to each respective rescaling of the horizontal axis. The choice of this rescaling, the thresholds indicated by the vertical lines in Fig. \ref{purity_fig}(b) and Fig. \ref{purity_fig}(c) and therefore the universal nature of the building of entanglement observed will become clearer in Section \ref{physical_mechanism}. Finally, we will refer to the small $T_u$ (or, equivalently, small $\Delta p$) regime as the \textit{Recoil entanglement regime} and the large $T_u$ (or, equivalently, large $\Delta p$) regime as the \textit{Doppler entanglement regime}, due  to the parameters that regulate these regimes.

\subsection{Schmidt rank}

For systems described by continuous variables, prepared in pure states, an useful entanglement measure is the so called \textit{Schmidt rank} $K$ \cite{Grobe1994,Ekert1998}, which accounts for the number of Schmidt modes in which a pure state can be decomposed. Thus the larger the value of $K$, the higher the amount of entanglement the system possesses. On the other hand, $K=1$ indicates that the system may be written in factorable form, i.e., separable. Moreover, the Schmidt rank is related to the purity of the reduced state of any of the parts of the system as
\begin{equation}
 K=\frac{1}{p_a}.
\end{equation}
We are going to use $K$ to interpret the nature of the entanglement for  the Recoil and Doppler regimes, as the number of Schmidt modes represents the number of independent product states for the atom-light system in which the entangled state can be decomposed. From now on, we will focus on the case of the emission of Cesium in the line $D_2$, which has the most well defined separation of the two high-entanglement regions as seen from Fig. \ref{purity_fig}.

%Results of the numerical calculation of the Schmidt rank for this case are presented in Fig. \ref{G2}. First of all note the presence of three separated domains: two opposite regions of high entanglement $(K>2)$, for both low and high $T_u$, and an approximately separable interval $(K\approx 1)$, for intermediate $T_u \sim \mathcal{O}(T_D)$. 

%\begin{figure} 
 % \centering
  %\includegraphics[width=9cm]{Schmidt_Cs-D2_(1).pdf}
  %\caption{Schmidt rank as a function of $T_u/T_R$ for the $\mathrm{Cs-D_2}$ Line. The solid green line represents the purity calculated from the small $\Delta p$ approximation [eq.\eqref{smallpapprox}], while the solid red line represents the purity calculated from the large $\Delta p$ approximation [eq.\eqref{largepapprox}]. The middle vertical line represents a minimum of entanglement where $T_u = T_D$. The shaded areas describe the range of $T_u$ where we obtain each entanglement regime, that is, $T_u \leq T_R$ for the Recoil entanglement regime and $T_u \geq 4 T_D^2/T_R$ for the Doppler entanglement regime. \color{red}{refazer figura trocando as energias pelas respectivas temperaturas. Deslocar legendas internas da figura para a esquerda, para não cortar a curva para $T_u \gg T_R$.}}
%  \label{G2}
%\end{figure}

\section{Physical considerations regarding the entanglement regimes} \label{physical_mechanism}

In this section we aim to provide physical intuition about the high entanglement regimes we found. Moreover, we intend to obtain a roughly accurate threshold for the high entanglement regimes (we use $K \geq 2$ as an indicator of high entanglement). The physical reasoning that will permeate this section is that as the effect of one subsystem upon the other becomes more and more measurable, more correlations are created in the system, leading to entanglement.

\subsection{Recoil entanglement}

For the small $\Delta p$ regime, the Recoil entanglement regime, we are dealing with an atom with a well defined initial momentum, $\vec{p}_i \approx 0$. This means that the spontaneous decay process behaves as a classical disintegration process, where one part of the system goes in a given direction while the other part goes in the opposite direction. While the momentum modulus of the atom or photon is well known (explicitly, $p_f \approx \hbar \omega_0/c$ with an uncertainty of $\hbar \Gamma/c$), the direction in which this process occurs is completely undetermined due to the nature of spontaneous emission. Using Eq.\eqref{Gen_State_a}, this yields a state of the form:
\begin{eqnarray}
    \ket{\psi} \approx \int \dd^3 k ~ C(0,\vec{k})\ket{-\hbar\vec{k}}_{at}\ket{\vec{k}}_{ph}\,,
\end{eqnarray}
which is a continuous non-separable state analogous to a (maximally entangled) Bell state. Note that: $k \in (k_0 - \Gamma/2c, k_0 + \Gamma/2c)$, which means that $k$ does not vary by a lot, while the angular variables range in the usual intervals. This means that the entanglement is in fact mostly encoded in the angular variables of the system. 

Finally, this discussion elucidates the fact that the Recoil entanglement arises in the regime where recoil effects are more relevant, that is, when the recoil due to the photon emission is big enough to change measurably the atomic initial momentum.  Therefore, a natural threshold for the Recoil entanglement is:
\begin{equation}\label{recoilthreshold}
    T_u = T_R\,,
\end{equation}
or, $\Delta p = \sqrt{2}\hbar \omega_0/c$, with the atomic momentum uncertainty being close to the photon momentum. 

\begin{figure}[ht!]
\centering
\includegraphics[width=0.5\textwidth]{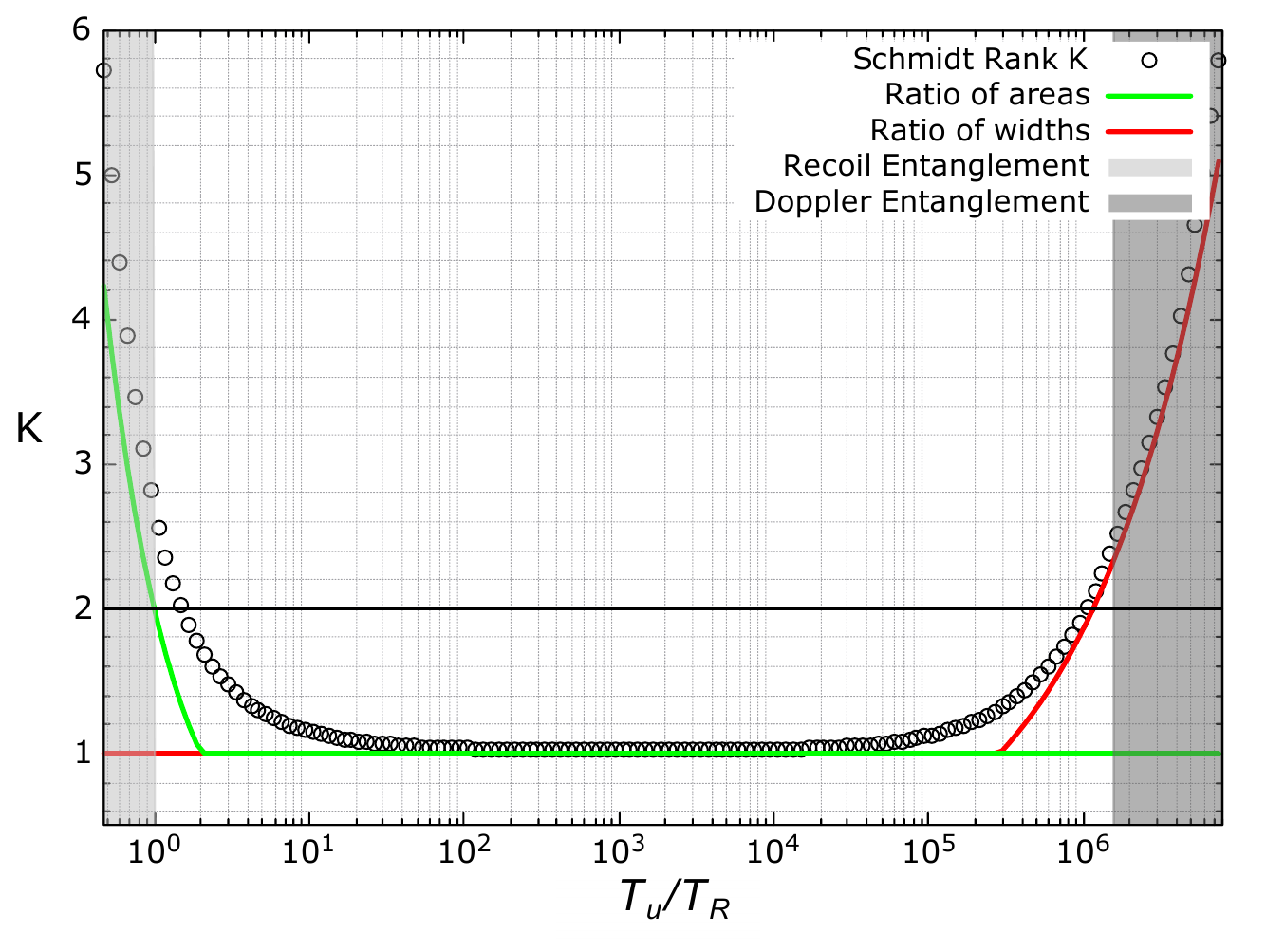}
\caption{\textmd{Comparison of the Schmidt rank calculated from estimates \eqref{ranksmallestimate} (solid green line) and \eqref{ranklargeestimate} (solid red line) with the actual Schmidt rank. Both estimates work better in the Recoil and Doppler entanglement regimes respectively.}}
\label{rankgammas}
%\par\medskip\ABNTEXfontereduzida\selectfont\textbf{Source: The author.}  \par\medskip
\end{figure}

The open circles in Fig. \ref{rankgammas} show the Schmidt rank K as a function of $T_u/T_R$ for the $\mathrm{D_2}$-Line of Cesium. As we can see from the light gray shaded area in the left part of the graph in Fig. \ref{rankgammas}, \eqref{recoilthreshold} gives a reasonable threshold for the Recoil entanglement regime, as $K \geq 2$. Moreover, if we rewrite \eqref{recoilthreshold} as $\frac{T_u}{T_R} = 1$, we obtain the threshold depicted in Fig. \ref{purity_fig}(b) and we can note that using this rescaling, the threshold becomes independent of physical parameters describing the spectral lines.  For uncertainty temperatures, $T_u$, greater than the recoil temperature, $T_R$, of the considered spectral line, we should not observe any considerable amount of entanglement due to this mechanism, since the atom momentum uncertainty becomes greater than the photon momentum and the correlation between the photon propagation direction and the final atom momentum decreases.

%, where Fig. \ref{rankgammas} corroborates with this statement.

We can understand the increase in the Schmidt rank as we decrease $T_u$ as follows: consider first a sphere of radius $\hbar \frac{\omega_0}{c}$ and, therefore, surface area of $4 \pi \cdot \hbar^2\frac{\omega_0^2}{c^2}$. Consider now a small circle living on the surface of the previous sphere with radius $\Delta p$, and spanning an area of $\pi \Delta p^2$. We argue that the Schmidt rank can be understood as counting the number of small circles that can live in the surface of the big sphere, that is:
\begin{eqnarray}\label{ranksmallestimate}
    K \approx \frac{4 \pi \cdot \hbar^2\frac{\omega_0^2}{c^2}}{\pi \Delta p^2} = 2 \frac{T_R}{T_u}\,.
\end{eqnarray}

Note that the counting we performed gives roughly the number of independent atomic scattering modes that fit in a sphere of noise with a radius comparable to the momentum recoil suffered by the atom. We can compare the estimate \eqref{ranksmallestimate} to the actual calculated Schmidt rank using Fig. \ref{rankgammas} and noting that the green solid line, representing the ratio of areas we defined, follows closely the scattered points for $T_u \leq T_R$. Note that we define $K = 1$ when: $2\, T_R/T_u \leq 1$.

%\begin{figure}[ht!]
%\centering

%\caption{\textmd{Linear entropy as a function of $T_{eff}/T_D$ for different atomic spectral lines.}}
%\label{figura12}
%\includegraphics[width=\textwidth]{images/chapter3/Linear_Entropy.png}

%\par\medskip\ABNTEXfontereduzida\selectfont\textbf{Source:% The author.}  \par\medskip
%\end{figure}

\subsection{Doppler entanglement}

In the Doppler entanglement regime, where the atom momentum uncertainty is much greater than the photon momentum, such that the photon emission almost do not change the atom momentum, we turn back to Eq.~\eqref{StateC}, to note that the effect of the recoil of the atom over the photon's central frequency is a blueshift or redshift depending on the emission direction in relation to the atom's final momentum direction:
\begin{eqnarray}
    \omega^0 - \omega_0 = \bigg(\frac{\omega_0}{mc}\bigg)\cdot q_{||}\,,
\end{eqnarray}
where $\omega^0$ is the new entanglement-induced emitted photon's central frequency and $q_{||} = \hat{k}\cdot \vec{q} =  q\, \cos \gamma_1\,$ is the component of the atom's momentum in the direction of the emitted photon, where $\gamma_1$ is the angle between $\hat{k}$ and $\vec{q}$ (Fig. \ref{G10}). Therefore, the central frequency of the emitted photon can be blueshifted if $q_{||} > 0$, or redshifted if $q_{||} < 0$, that is, if the momentum of the atom is aligned parallel to the direction of the emitted photon or anti-parallel to it. 

We should then expect an increase in entanglement as the shift in frequency becomes greater than the natural linewidth $\Gamma$. Considering that the components that most participate in the effect are of the order of $\Delta p$, that is, $q_{||} \approx \Delta p$, we can write the condition for entanglement as:
\begin{eqnarray}
    \bigg(\frac{\omega_0}{mc}\bigg)\, \Delta p \geq \Gamma\,,
\end{eqnarray}
or equivalently, solving for $\Delta p$ and writing the solution in terms of $T_u$, $T_R$, and $T_D$, we obtain the Doppler threshold:
\begin{eqnarray}\label{dopplerthreshold}
    T_u \geq 4 \bigg(\frac{T_D}{T_R}\bigg)T_D = T_{DE}\,,
\end{eqnarray}
which defines the \textit{Doppler entanglement temperature}. Note that $T_u = T_{DE}$ is precisely the threshold being depicted in Fig. \ref{purity_fig} (c), and just like in the Recoil entanglement regime, if we write $\frac{T_u}{T_{DE}} = 1$, this threshold becomes independent of physical parameters. As we can see from Fig. \ref{rankgammas}, the shaded dark gray area in the right part of the graph [which illustrates the threshold \eqref{dopplerthreshold}] represents well the region where $K \geq 2$. An analog view of this effect is the following: we can understand the threshold \eqref{dopplerthreshold} as the minimum uncertainty temperature such that the (homogeneous) Doppler broadening induced by it is greater than the natural linewidth of the spectral line. Explicitly:
\begin{equation}\label{effectivegamma}
    \Gamma_e = \sqrt{\frac{k_B T_u}{mc^2}}\,\omega_0 \geq \Gamma,
\end{equation}
where we define the \textit{effective natural linewidth}, $\Gamma_e$, and now, eq.\eqref{effectivegamma} can be rewritten as eq.\eqref{dopplerthreshold}. Note that we use the term \textit{homogeneous} to emphasize the fact that we are working with a single atom and making parallels with an atomic ensemble at an effective temperature $T_u$.

We can then understand the increase in the Schmidt rank as a quantification of how many $\Gamma$'s fit in a given $\Gamma_e$. More accurately, since the Gaussian and Lorentzian distribution are of different natures, we compare the Full-Width at Half Maximum (FWHM) of the Doppler induced spectrum $\sqrt{8 \log 2}\, \Gamma_e$ to $\Gamma$, and argue that:
%since the Lorentzian distribution has an undefined variance, we should not compare directly the Doppler's standard deviation $\Gamma_e$ to $\Gamma$, though we can define the Full-Width at Half Maximum (FWHM) of the Doppler induced spectrum as $\sqrt{8 \log 2}\, \Gamma_e$ and argue that:
\begin{eqnarray}\label{ranklargeestimate}
    K &=& \sqrt{8 \log 2}\,\frac{\Gamma_e}{\Gamma} = \sqrt{\frac{8 \log 2\,\omega_0^2}{\Gamma^2 m c^2}} \cdot \sqrt{k_B T_u} \nonumber \\
    &=& \sqrt{2 \log 2}\frac{\sqrt{T_R}}{T_D} \sqrt{T_u} \,.
\end{eqnarray}

In fact, as we can see from Fig. \ref{rankgammas}, the red solid line, representing the ratio of widths we just defined, follows along the calculated Schmidt rank specially as $K$ increases from $2$, that is, in the Doppler regime. Note that we define $K = 1$ when: $\sqrt{8 \log 2}\, \Gamma_e < \Gamma$.

\begin{figure}[ht!]
\centering
\includegraphics[width=0.5\textwidth]{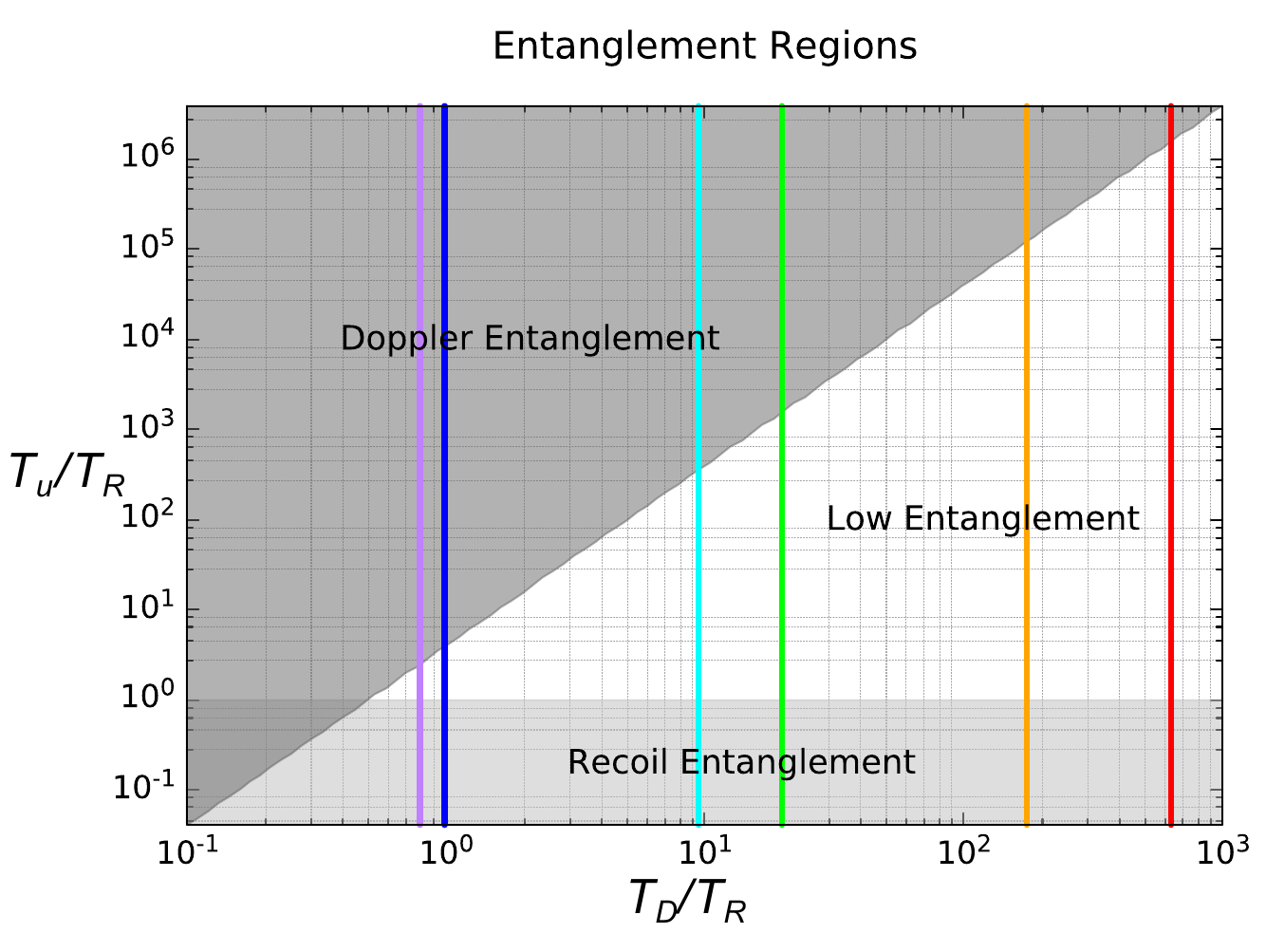}
\caption{\textmd{Phase diagram elucidating all observed high entanglement regimes and low entanglement regions. The vertical lines represent different spectral lines (color coded with Fig. \ref{purity_fig}) labeled by different values of $T_D/T_R$. Note that the thresholds we obtained in Eq. \eqref{recoilthreshold} and Eq. \eqref{dopplerthreshold} work well to estimate the $K \approx 2$ regions only for cases far enough away from the mixed entanglement regime.}}
\label{phasediagram}
%\par\medskip\ABNTEXfontereduzida\selectfont\textbf{Source: The author.}  \par\medskip
\end{figure}

Finally, we can summarize the thresholds we obtained in Fig. \ref{phasediagram} and look at the results shown in Fig. \ref{purity_fig} in an analogous manner: note that each spectral line we considered presents a different value of $T_D/T_R$, which we represent as the vertical lines in Fig. \ref{phasediagram}. We showed that the interplay between Doppler and Recoil energies plays a crucial role in the observed entanglement behavior, and following along a given spectral line shown in Fig. \ref{phasediagram} we can observe the transitions from Recoil entanglement to a low entanglement plateau to Doppler entanglement as is the case for the $\mathrm{Cs-D_2}$ line. As for the Strontium narrow line case, we can see that the system does not leave considerably any region of high entanglement, staying roughly highly entangled at all considered values of $T_u$. In fact, we can note that for certain values of $T_u$ and $T_D/T_R$, the effects of the Recoil and Doppler regimes of entanglement overlap, and no source of entanglement can be separated from the other. Explicitly, if $T_D/T_R \leq \frac{1}{2}$, we can expect our system to be highly entangled for all values of $T_u$.

% , either on the Doppler regime $(T_u > T_R)$ or in the mixed entanglement regime $(T_u < T_R)$.

\section{Conclusions and perspectives}
In this paper we reviewed the problem of spontaneous emission using the formalism developed by \cite{Weisskopf1930} and \cite{Rzazewski1992}, considering external atomic degrees of freedom, and quantified the entanglement encoded in the continuous variables of the system through a calculation of the purity of the reduced atomic state.

By doing so, we observed regimes of high entanglement usually separated by a low entanglement region. Depending on the considered spectral line, this low entanglement region may express itself as a plateau, a peak of purity higher than $P_a = 0.5$ (yielding a Schmidt rank lower than $K = 2$) or a peak that does not cross the $P_a = 0.5$ threshold (not characterizing a true low entanglement region). 

Due to their nature, we defined the Recoil and Doppler regimes of entanglement and obtained thresholds in $T_u$ which need to be crossed in order to achieve these entanglement regimes. We also defined a third regime where both regimes are active and yields atom-photon systems that are always highly entangled. We observed that, physically, recoil effects play a crucial role in the Recoil entanglement regime, while homogeneous Doppler shifts play a crucial role in the Doppler regime of entanglement, that is, these effects are responsible for the creation of quantum correlations in the atom-photon system. We also presented physical considerations that could describe the increase of the Schmidt rank supported by estimates that agree qualitatively with the calculated Schmidt ranks.

% Finally, we investigated the role of entanglement in the discrimination of two physically distinct quantum states and observed that the presence of entanglement in the system allows us to better distinguish between different quantum states that arise naturally in the process of spontaneous decay. This result can be useful for quantum discrimination protocols as well as for quantum imaging. 

Further generalizations of our model and method of quantification of entanglement are in order. Namely, extending the investigation to an atomic ensemble in thermal equilibrium with a reservoir at temperature $T$ spontaneously emitting a single photon (and therefore creating a Dicke state) should provide a theoretical model that is easier to implement experimentally. Note that this extension allows a further generalization regarding the spontaneous emission of $n$ photons, generating, in principle, multipartite entanglement, which would lead to the necessity of a more robust entanglement quantification. Though more complex, the problem of $m$ atoms spontaneously emitting $n$ photons is also of importance in the construction of quantum protocols.

\section{Acknowledgments}
This work was supported by the Brazilian funding agencies Conselho Nacional de Desenvolvimento
Cient\'ifico e Tecnol\'ogico (CNPq - INCT Grant No.  $465469/2014-0$), Coordena\c{c}\~ao de Aperfei\c{c}oamento de Pessoal de N\'ivel Superior (CAPES - PROEX Grant No. $23038.003069/2022-87$), Funda\c{c}\~ao de Amparo \`a Ci\^encia e Tecnologia do Estado de Pernambuco (FACEPE), and Funda\c{c}\~ao de Amparo \`a Pesquisa do Estado de S\~ao Paulo (FAPESP Grant No. $2021/06535-0$). J.C.C.C. thanks the Office of Naval Research (ONR Grant No. $\mathrm{N}62909-23-1-2014$). A. F. acknowledges funding by Funda\c{c}\~ao de Amparo \`a Ci\^encia e Tecnologia do Estado de Pernambuco - FACEPE, through processes $\mathrm{BFP}-0168-1.05/19$ and $\mathrm{BFP}-0115-1.05/21$. J. C. C. C. and A. F. contributed equally to this work.

\end{document}